\documentclass[sigconf]{acmart}
\AtBeginDocument{%
  }

\copyrightyear{2026}
\acmYear{2026}
\setcopyright{cc}
\setcctype{by-nc-nd}
\acmConference[CHIWORK '26]{Proceedings of the 5th Annual Symposium on Human-Computer Interaction for Work}{June 22--25, 2026}{Linz, Austria}
\acmBooktitle{Proceedings of the 5th Annual Symposium on Human-Computer Interaction for Work (CHIWORK '26), June 22--25, 2026, Linz, Austria}
\acmDOI{10.1145/3808045.3808063}
\acmISBN{979-8-4007-2429-9/2026/06}

\usepackage{pgfplots}
\usepackage{pgfplotstable}
\pgfplotsset{compat=1.18}
\usetikzlibrary{patterns}
\usepackage{tcolorbox}
\usepackage{multirow}

\begin{document}

\title{Cheap Expertise: Mapping and Challenging Industry Perspectives in the Expert Data Gig Economy}

\author{Robert Wolfe}
\affiliation{
  \institution{Rutgers University}
  \city{New Brunswick, NJ}
  \country{United States}}
\email{robert.wolfe[at]rutgers.edu}

\author{Aayushi Dangol}
\affiliation{
  \institution{University of Washington}
  \city{Seattle, WA}
  \country{United States}}
\email{adango[at]uw.edu}

\renewcommand{\shortauthors}{Wolfe \& Dangol}

\begin{abstract}
Demand for expert-annotated data on the part of leading AI labs has created an expert gig economy with the potential to reshape white collar work and society's understanding of expertise. In this research, we study the vision for the future of expertise described in the public communication of five industry data annotation organizations and their CEOs, as reflected on social media feeds and public appearances on podcasts. We find that the industry envisions AI expertise as cheap, meaning that it can offer a better return on investment than human expertise. Human expertise, meanwhile, is viewed as an extractable resource, the value of which can be judged relative to AI expertise. Finally, institutional expertise (such as that created or possessed by universities and corporations) is viewed as in need of liberation or reform, such that it can be incorporated into the latest artificial intelligence systems. Our findings have implications for human experts, whose professional lives may be transformed and revalued by this industry, as well as for societal institutions that mediate expertise. We close this work with a series of provocations intended to elicit consideration of how society can best approach an AI-driven expert gig economy and the cheap expertise it intends to produce.
\end{abstract}



\maketitle

\section{Introduction}

In June 2025, Meta acquired 49\% of the industry-dominant AI data labeling company Scale AI at a cost of \$14.8 billion, and assigned its CEO to head the newly founded Meta SuperIntelligence division \cite{perrigo2025scale}. In the wake of the transaction, leading industry AI labs were reported to have reallocated data labeling contracts to rival organizations, uncertain of whether Meta's investment would allow the company to observe their strategy for training frontier AI models \cite{perrigo2025scale,nieva2025mercor}. While some of these competing organizations were well-established in the field, like Surge AI \cite{conrad2025surge}, others appeared to rebrand almost overnight. The university career networking company Handshake, for example, made announcements about its Handshake AI platform days after the Scale AI transaction, and its CEO reported it had tripled its contracts after rebranding \cite{perrigo2025scale,konrad2025handshake}.

In the months that followed, these organizations attempted to position themselves as the best-suited to provide AI labs like OpenAI and Anthropic with a scarce resource: human expertise. Specifically, they sought to provide expert-authored demonstrations of how to perform tasks requiring knowledge or capability rarefied enough that it had not yet been encoded in the leading AI models. To do so, they built and expanded an expert gig economy offering (often lucrative) hourly wages to domain experts \cite{nieva2025mercor}, creating platforms intended to reshape white collar knowledge work and society's shared understanding of expertise.

The present work seeks to describe the vision for the future of expertise offered by the most influential organizations operating in the expert gig economy. To do so, we study the public communication of five organizations (Surge AI, Handshake AI, Turing, Scale AI, and Mercor), decomposing their vision into three complementary aspects. Specifically, we analyze the industry's perspective on \textbf{1) AI expertise}, meaning the expert knowledge and capability encoded in machines via demonstrations labeled by humans; \textbf{2) human expertise}, meaning the expert knowledge and capability achieved by individuals as a result of learning and lived experience; and \textbf{3) institutional expertise}, meaning the expert knowledge and capability produced and legitimized by societal institutions like corporations and universities.

To understand this vision, we employ a directed content analysis methodology \cite{assarroudi2018directed}, first reviewing all content posted publicly to the X platform feeds of the organizations and their CEOs between June and December of 2025. Following an inductive-deductive coding approach \cite{fereday2006demonstrating}, we produce a codebook with inductive subcodes organized under three primary codes for the three forms of expertise (AI, Human, and Institutional). We then apply these codes to the transcripts of 29 podcasts on which the CEOs of the organizations appeared, amounting to more than 20 hours of audiovisual content. Following a thematic analysis approach \cite{braun2019reflecting}, and drawing on our coded social media and podcast data, we generate nine themes to characterize the industry's vision for the future of expertise, as highlighted in bold:

\begin{enumerate}
    \item \textbf{AI Expertise}: The industry presents AI expertise as \textbf{1) cheap}, meaning that it must generate a return on investment with more favorable margins than one could attain by purchasing the services of a human expert; \textbf{2) in need of recontextualization}, meaning that AI expertise must be obtained and tested in scenarios ever more faithful to the real world, driving the development of evaluations and learning environments by the firms themselves; and \textbf{3) unrealized}, and unlikely to be realized in the very near future, meaning that the expert gig economy will remain necessary to produce specialized forms of AI expertise for years to come.
    \item \textbf{Human Expertise}: The industry presents human expertise as \textbf{1) decoupled from the expert}, meaning that human expert knowledge can be viewed as an extractable resource to be harvested, rather than necessarily tied to the lived experience of the individual expert; \textbf{2) relative to AI expertise}, meaning that the degree of knowledge or capability possessed by an individual can be judged by whether it has already been encoded in a machine; and \textbf{3) elastic}, meaning that it is contingent on the present needs of the AI labs, and has value where it can be efficiently harvested to respond to those needs.
    \item \textbf{Institutional Expertise}: The industry presents institutional expertise as \textbf{1) in need of liberation}, meaning that institutions' incentive to protect their expert knowledge and capability prevents the advancement of AI expertise, such that the expert gig economy offers a way to extract or approximate that otherwise unattainable resource; \textbf{2) performative}, meaning that the credentials and social signals of expertise provided by traditional institutions offer too little real-world value in the age of AI, such that the act of AI data annotation can serve as a more reliable measure of capability; and \textbf{3) inverted} by AI augmentation, meaning that traditional institutional hierarchies based on achievement and experience must be upended by the availability of AI expertise, empowering instead those who most fully integrate AI into their professional lives.
\end{enumerate}

\noindent Put succinctly, the data annotation industry presents AI expertise as cheap but unrealized, human expertise as measurable and extractable, and institutional expertise as in need of liberation and reform. These findings reflect the priorities and financial motives of the data annotation industry, and to some degree they may not seem surprising. Yet they also suggest the potential for a markedly different future for knowledge work, and one that society may need to swiftly reckon with, in that the data annotation firms now command substantial capital to realize their vision and already employ millions of knowledge workers to provide annotations on an hourly basis. Much prior work has charted the social and economic precarity of workers who work in algorithmically mediated gig economies \cite{wood2019good,wood2021platform,alacovska2025algorithmic}; our research charts a future that could extend such insecurity to a wide array of white collar knowledge workers, many of whom have committed considerable time and financial resources to obtain their credentials. We thus assert the need for critical conversations that address this potentially destabilizing future. In addition to a discussion that suggests directions for future work, we close this research with a series of provocations intended to prompt consideration of how society can respond to the expert gig economy and the cheap expertise it intends to produce.

\section{Related Work}

We review the related work on what expertise is and how it is evolving; on the development of gig economies and the precarity experienced by gig workers; and on AI data work, with specific attention to domain expert data annotation.

\subsection{Expertise}

Expertise is studied from many different perspectives across the social sciences. Speaking from a psychological perspective, 
\citet{bourne2014expertise} discuss the expert as defined by a combination of both knowledge and capability, asserting that ``to describe expertise is to identify the endowed resources, catalog the knowledge, and specify the skills of a person who is capable of performing in some domain at the very highest level, achieved by few others.'' Offering a sociological account that places an expert in the context of the individuals and society they serve, \citet{grundmann2017problem} describe expertise as ``relational in a double sense: it relates to clients and it relates to their needs, which often is the need for guidance in decision-making.'' From an anthropological perspective, \citet{carr2010enactments} describe expertise as enacted by the expert, and defined by ``socialization practices[\dots]; evaluation[\dots]; institutionalization[\dots]; and naturalization, or the essentialization of expert enactments as bodies of knowledge.'' Our understanding of expertise in the present work refers primarily to the psychological perspective that views expertise as a combination of rarefied knowledge and ability. However, our work is also informed by sociological and anthropological perspectives, including in its description of the industry's vision for institutional expertise.

Recent work suggests that advances in AI will substantially alter the work of experts and the perception of expertise in society. \citet{susskind2017future} argue that future technologies, most notably AI, will augment and ultimately automate expert professions including law, accounting, and medicine, rendering expertise more accessible and affordable, but also altering and revaluing the channels through which expertise is obtained and distributed. \citet{eloundou2024gpts} assert that generative AI \textit{is} such a general-purpose technology, especially when paired with co-invented systems that extend its functionality. Subsequent research finds that generative AI can serve numerous functions in knowledge work, ranging from meeting planning \cite{scott2025does,chen2025we}, to coordinating group work \cite{he2024ai,ju2025collaborating}, to providing automated feedback to individuals and teams \cite{almutairi2025taifa,hsu2025helping}, among many others \cite{kobiella2025efficiency,das2024teacher,muller2025exploration,drosos2025makes}, with applications in fields including speech pathology \cite{suh2024opportunities,lewis2025exploring}, air traffic control \cite{schon2025cleared}, software engineering \cite{meem2025investigating,feldman2024non}, data integration \cite{han2025can,kachari2026integrating}, professional fact checking and journalism \cite{hollanek2025ai,wolfe2024implications,wolfe2024impact,das2023state,cools2024uses}, and education \cite{dangol2025relief,ammari2025students}. Accordingly, \citet{constantinides2025future} contend that AI has already thoroughly infiltrated workplaces, such that the future of work should be seen as ``blended'' in the sense that it is shared with AI co-actors and mediators, rendering human output inseparable from the algorithms that shape it. Our research builds on this line of work by studying the role of the expert gig economy in shaping the future of expertise.

\subsection{Gig Economies}

Our research studies a ``gig economy'' intended to extract expert data annotation to improve large language models. Following prior work, we define a gig economy as a market for independent contracting work facilitated via an electronic interface, usually a platform or app built for communication and management of gig workers \cite{woodcock2020gig, srnicek2017platform}. Though freelance and independent contracting work long precede the idea of app or platform mediated employment \cite{kazi2014freelancer,van2013drivers}, a ``gig'' economy in particular is sociotechnical, such that a social subsystem for employment shapes and is mutually shaped by a technical subsystem provided by platforms \cite{dedema2024socio,ackerman2000intellectual,chopra2018sociotechnical}.

Much prior work studies how gig economies transform sectors of employment that require less formal training, such as driving for rideshares or performing domestic tasks \cite{berger2019uber,heiland2021controlling}. However, knowledge work has also seen the rise of gig economies prior to the current market for expert annotations, most notably in the form of freelancing platforms like Upwork and Fiverr that post contracts for information economy jobs, like building a website or optimizing content for a search engine \cite{blyth2024self,fulker2024cooperation}. While these markets for digital freelance work anticipate the expert gig work we study in this research, they are themselves reported to be declining due to the advent of general-purpose AI, as \citet{demirci2025ai} find evidence for a 21\% decline in automation-prone job listings in the eight months after the release of ChatGPT.

This ephemeral property of the market for skilled online work also points to the vulnerability of many gig workers. \citet{wood2021platform} use the term ``algorithmic insecurity'' to refer to workers' experiences navigating opaque and often unstable gig markets, typified by arbitrary or unexplained shifts in one's reputation and associated earning capacity. \citet{graham2019global} locate the effects of the global online gig economy in a ``planetary labor market'' that ultimately undermines the structural and associative power of digital workers. In studies of gig knowledge work, \citet{wilkins2022gigified} find that freelance knowledge workers are able to draw on a more limited pool of resources and connections than in-house employees, while \citet{alacovska2025algorithmic} find that digital gig workers often report experiencing mistrust, anxiety, and fear due to abusive algorithmic management. Moreover, \citet{lascuau2024sometimes} find that gig workers' temporal flexibility, often seen as a benefit of the job, is more limited than might be expected by virtue of having to be on call for opportunities. Gig work is nonetheless appealing to some individuals, as it can offer variety, flexibility, and autonomy not available in full-time jobs \cite{wood2019good,myhill2023job}, and studies show that gig knowledge workers often prioritize finding reliable, quality work that promotes self-growth \cite{kim2025decoding}.

\subsection{AI Data Work}

AI systems require human-labeled data for ``supervised learning,'' referring to phases of training that learn from explicit labels rather than patterns in data \cite{nasteski2017overview}. After modern chatbots and agents are ``pretrained'' in an unsupervised manner on large text corpora \cite{brown2020language,raffel2020exploring}, they undergo a ``post-training'' phase that uses human labels or demonstrations to align them with specific ethical norms and equip them with specialized knowledge and capabilities \cite{chu2025sft,ouyang2022training,ke2025survey}. It is this post-training phase that produces demand for a wide variety of human data inputs \cite{muldoon2024typology,attard2025ethics}. Where alignment work tends to involve systematic evaluation of LLM adherence to a policy and is often outsourced to the global south \cite{perrigo2023openai,bai2022training,okolo2024moving}, the data to improve AI agents and ``reasoning'' models can involve more complex demonstrations, of the sort that only a domain expert with specific knowledge and experience in a field can provide \cite{reddy2025towards}. For example, improving the performance of a model on mathematical reasoning might involve the curation of step-by-step examples of problem solving, sometimes referred to as ``reasoning traces,'' created by an expert mathematician \cite{kumar2025llm}; improving a model's ability to carry out tasks online might similarly involve observing visual data of a capable internet user carrying out such tasks via a computer use interface \cite{sager2026comprehensive,guo2026susbench}.

Despite the centrality of domain experts for obtaining appropriate AI data, prior work observes a tenuous and reductive relationship between organizations and the expert annotators they employ. \citet{sambasivan2022deskilling} find that AI developers viewed domain expert workers as non-essential, lazy, and corrupt, in need of better surveillance or gamification strategies to collect better data. \citet{rothschild2024problems} similarly find that data requesters on crowd platforms hold simplistic views of worker identities, based more on performance on qualification tasks than on more contextualized measures of expertise. Work by \citet{sun2022comparing} demonstrates that, while experts provided better quality labels on a task in their domain, they also found the work of data annotation more tedious and repetitive, suggesting negative consequences for expert job satisfaction. Recent work also finds that experts may experience skill degradation and reduction in work opportunities due to technologies produced from expert data labor \cite{natali2025ai,ferdman2025ai,van2025unintended}.

Finally, recent work also points to shortcomings in the body of research on expert data work. \citet{diaz2024makes} find that much research on expertise in the development of AI systems leaves the term ``expertise'' undefined, or instead defines it simplistically as ``discrete information that can be memorized and reproduced relatively easily, such as through data annotation,'' leaving the social components of expertise unaddressed, and leaving little space for marginalized or non-canonical forms of expertise. Moreover, highlighting the centrality of ``cost reduction'' that our notion of cheap expertise also addresses, \citet{constantinides2025ai} contend that automation-centered paradigms dominate discussions of the AI-driven future of work, marginalizing human-centered AI programs that prioritize collaboration. We build on this research by describing the industry's vision to extract expertise and reproduce a cheap form of it in AI.

\begin{table*}[]
    \centering
    \begin{tabular}{l|c|c|c}
    \toprule
    Organization / CEO  & X Posts & Podcast Appearances & Podcast Duration  \\
    \midrule
    \textbf{Surge AI} & 29 & --- & 00:00:00 \\
    Edwin Chen (Surge AI CEO) & 42 & 5 & 04:32:42 \\
    \midrule
    \textbf{Scale AI} & 82 & --- & 00:00:00 \\
    Jason Droege (Scale AI CEO) & 10 & 3 & 01:59:07 \\
    \midrule
    \textbf{Handshake AI} & 75 & --- & 00:00:00 \\
    Garrett Lord (Handshake AI CEO) & 88 & 8 & 05:28:34 \\
    \midrule
    \textbf{Mercor} & 43 & --- & 00:00:00 \\
    Brendan Foody (Mercor CEO) & 77 & 7 & 04:09:05 \\
    \midrule
    \textbf{Turing} & 519 & --- & 00:00:00\\
    Jonathan Siddharth (Turing CEO) & 289 & 6 & 04:37:46 \\
    \midrule
    \textbf{Total} & 1,254 & 29 & 20:47:14 \\
    \bottomrule
    \end{tabular}
    \caption{We use a directed content analysis methodology, first studying X posts by five organizations and their CEOs in the expert data annotation industry, and then applying our codebook to 29 podcast appearances made by the CEOs.}
    \label{tab:datareviewed}
\end{table*}

\section{Data}

This research studies the public communication of five organizations in the expert data annotation industry after the June 2025 investment of Meta in Scale AI. The data considered falls into two domains: social media posts shared by the organizations or their CEOs to their public feeds on the X platform during our data collection period, and audio or video of podcast appearances made by the CEOs linked from those posts. Where an X post links to external content or includes a video clip in the post, we also follow the link and read the external content or watch the video included in the post in order to provide additional context for our analysis. Table \ref{tab:datareviewed} describes the data collected and reviewed, broken down by organization and CEO. As is clear from the table, Turing and Siddharth pursue a more aggressive social media communication strategy than the other organizations, though much of the content includes reposts and retweets.\footnote{We are unable to navigate to the beginning of the Turing feed, as the feed stops loading posts prior to June 25, 2025, a behavior we observe in several browsers. We believe that this could be due to the length of the feed. We do not believe that this materially affects our analysis, given the large amount of posts from Turing and Siddharth.} For podcasts, we report the number of podcast appearances by each organization's CEO, as well as the total duration of those appearances. Note that our analysis excludes internally produced topical podcasts not directly concerned with expert data annotation and not attended by the CEOs, such as the Human in the Loop podcasts produced by Scale AI.

\subsection{Organizations}
We select the organizations studied in this research by referring to two influential media articles surveying the leading organizations providing expert data labeling services after Meta's Scale AI investment. The first is a June 16, 2025 TIME Magazine article\footnote{See \url{https://time.com/7294699/meta-scale-ai-data-industry/}.} (identifying Handshake AI, Turing, and Scale AI), and the second is a September 3, 2025 Forbes Magazine article\footnote{See \url{https://www.forbes.com/sites/richardnieva/2025/09/03/ais-next-job-recruiting-people-to-train-more-ai/}.} (identifying Mercor, Surge AI, and Turing). Note that we exclude two organizations, Appen (mentioned by TIME) and Invisible Technologies (by Forbes), due to a lack of public communication on these organizations' public X feeds. We provide an overview of the five included organizations:

\begin{itemize}
    \item \textbf{Surge AI}: A data labeling firm founded in 2016 by CEO Edwin Chen, reported by Forbes to be valued around \$25 billion as of September 2025 \cite{nieva2025mercor}. Unlike most of its competitors, Surge AI is profitable, has never raised outside investment (though it may be in the process of doing so now), and is reported to bring in more than \$1 billion yearly in revenue \cite{conrad2025surge}. Surge AI describes its mission as contributing to artificial general intelligence that is not optimized for engagement but imbued with the richness of human experience \cite{surge2026extraordinary}.
    \item \textbf{Scale AI}: A data labeling firm founded in 2016 by former CEO Alexandr Wang and a co-founder \cite{katje2025wang}. After Wang's departure to Meta following its more than \$14 billion (49\%) investment in Scale AI \cite{perrigo2025scale}, current CEO Jason Droege assumed leadership \cite{capoot2025wang}. Though TIME reports OpenAI and Google reduced operations after the Meta investment \cite{perrigo2025scale}, Scale AI officials assert the independence of its data labeling business from Meta \cite{nieva2025mercor}.
    \item \textbf{Handshake AI}: A data labeling branch of Handshake, an organization founded in 2014 by CEO Garrett Lord and two co-founders that primarily offers an employment portal for universities \cite{yeung2016handshake}. Where Handshake offers an employment platform, Handshake AI offers a gig platform that mediates between current and former university students and AI labs seeking data annotation services \cite{handshake2026handshakeai}. Though Handshake maintains the importance of its career portal business, it also laid off about 15\% of its workforce, working in the career portal arm of the business, when it announced Handshake AI \cite{konrad2025handshake}. Handshake was valued at \$3.5 billion in 2022 \cite{roof2022handshake}.
    \item \textbf{Mercor}: A recruiting organization founded in 2023 by CEO Brendan Foody and two co-founders, with an initial focus on creating an AI-driven platform to match job seekers to interested firms \cite{nieva2025mercor}. Though Forbes reports that Foody still views this as the firm's primary focus, data labeling is now a core part of the business \cite{nieva2025mercor}, and a central reason Mercor is valued at \$10 billion as of October 2025 \cite{iyer2025mercor}.
    \item \textbf{Turing}: A data labeling firm led by CEO Jonathan Siddharth and valued at \$2.2 billion as of March 2025 \cite{lunden2025turing}. TechCrunch reports that Turing's business initially centered around management tools for remote teams, but pivoted to expert data annotation around 2022 \cite{lunden2025turing}. Its leadership often refers to Turing as a research accelerator working in partnership with organizations trying to improve or implement AI \cite{turing2026company}.
\end{itemize}

\subsection{Data Collection Period}

To ensure we obtain a quantity of data tractable for in-depth qualitative study, we focus our analysis on content posted between June 1, 2025 and December 31, 2025. The start date of the analysis corresponds both to the month Scale AI announced Meta's investment in the company \cite{perrigo2025scale} and to a significant increase in public communication by the data annotation firms, as they seek to position themselves to capitalize on a market thought to be seeking alternatives to Scale AI. For example, Surge AI CEO Edwin Chen makes his first ever podcast appearance on July 22, 2025. We choose to end the data collection period at the end of 2025 to afford time for qualitative analysis.

\begin{table*}[htbp]
\centering
\begin{tabular}{lllll}
\toprule
\textbf{CEO} & \textbf{ID} & \textbf{Date} & \textbf{Podcast} & \textbf{Duration} \\
\midrule
\multirow{5}{*}{Edwin Chen (Surge AI)} 
& T1 & 7/21/2025 & The Twenty Minute VC & 01:07:30 \\
& T2 & 7/24/2025 & No Priors & 00:32:38 \\
& T3 & 9/16/2025 & Gradient Dissent & 00:54:30 \\
& T4 & 12/07/2025 & Lenny's Podcast & 01:10:05 \\
& T5 & 12/15/2025 & Unsupervised Learning & 00:47:59 \\
\midrule
\multirow{3}{*}{Jason Droege (Scale AI)} 
& T6 & 9/5/2025 & Technology Business Programming Network & 00:23:18 \\
& T7 & 9/17/2025 & Axios AI Summit & 00:12:14 \\
& T8 & 10/9/2025 & Lenny's Podcast & 01:23:35 \\
\midrule
\multirow{8}{*}{Garrett Lord (Handshake AI)} 
& T9 & 6/19/2025 & Technology Business Programming Network & 00:11:11 \\
& T10 & 6/18/2025 & Let There Be Light (ASU+GSV) & 00:51:47 \\
& T11 & 6/19/2025 & The Twenty Minute VC & 00:15:05 \\
& T12 & 6/26/2025 & Deirdre Bosa Live & 00:14:48 \\
& T13 & 7/21/2025 & Grit Podcast & 01:01:40 \\
& T14 & 7/24/2025 & The Peel & 01:35:40 \\
& T15 & 8/20/2025 & The Information TV & 00:09:00 \\
& T16 & 8/24/2025 & Lenny's Podcast & 01:09:23 \\
\midrule
\multirow{7}{*}{Brendan Foody (Mercor)} 
& T17 & 6/4/2025 & Unsupervised Learning & 00:43:53 \\
& T18 & 8/26/2025 & First Block & 00:26:35 \\
& T19 & 9/15/2025 & The Twenty Minute VC & 01:00:25 \\
& T20 & 9/18/2025 & Lenny's Podcast & 01:06:41 \\
& T21 & 9/18/2025 & Technology Business Programming Network & 00:16:04 \\
& T22 & 10/27/2025 & Technology Business Programming Network & 00:10:33 \\
& T23 & 11/5/2025 & TechCrunch Disrupt & 00:24:54 \\
\midrule
\multirow{6}{*}{Jonathan Siddharth (Turing)} 
& T24 & 7/14/2025 & ICML Expo & 00:59:38 \\
& T25 & 8/1/2025 & This Week in Startups & 00:20:24 \\
& T26 & 9/15/2025 & B2BaCEO & 01:02:14 \\
& T27 & 9/17/2025 & The Information TV & 00:09:03 \\
& T28 & 10/10/2025 & Sourcery & 00:49:33 \\
& T29 & 12/1/2025 & The Twenty Minute VC & 01:16:54 \\
\bottomrule
\end{tabular}
\caption{Podcast appearances made by the CEOs, with the transcript ID we use to reference them included in the ID column.}
\label{tab:podcasts}
\end{table*}

\subsection{Podcasts}

As shown in Table \ref{tab:podcasts}, the podcasts on which the CEOs appear can be characterized, for the most part, as channels for communication between technology startups and potential investors. The most popular podcasts for CEOs to appear on include the Technology Business Programming Network or TBPN (Jason Droege, Garrett Lord, and two appearances by Brendan Foody); The Twenty Minute VC (Venture Capitalist) or 20VC (Edwin Chen, Garrett Lord, Brendan Foody, and Jonathan Siddharth); and Lenny's podcast, a program discussing technology products and investment (Edwin Chen, Jason Droege, Garrett Lord, Brendan Foody). We thus note that much of the content studied in this research is intended for an audience interested in investing in tech organizations.\footnote{Note that we include Siddharth's appearance at the ICML Expo, Foody's appearance at TechCrunch Disrupt, and Droege's appearance at the Axios AI Summit, as these constitute public communication by the CEOs, though they are not strictly speaking podcasts. We decided to include this content because, with the exception of Siddharth's ICML talk, these appearances are conversational, similar to the podcasts we study; they are shared to institutional social media feeds; and the communication is primarily targeted to investors, as in the podcasts.} We believe that the statements made in these public forums provide direct and revealing insights into the objectives of the data annotation firms, though we must take care to also situate such statements in their original context.

\section{Methods}

To answer our primary research questions, we conduct a qualitative analysis of the data we collected, beginning with three deductive codes corresponding directly to our RQs: AI Expertise; Human Expertise; and Institutional Expertise. Our goal is to produce a set of themes that characterizes how the data annotation industry represents and intends to shape each of these forms of expertise. To do so, we employ a directed content analysis methodology \cite{assarroudi2018directed} and an inductive-deductive coding approach \cite{fereday2006demonstrating}, first applying our deductive codes to the threads collected from X, and adding more specific inductive codes within each top-level deductive code. After this initial round of coding, we hold a meeting to produce a codebook, which includes the three top-level codes and twenty-nine subcodes within those codes. In this meeting, we create definitions and inclusion criteria, adding representative posts to illustrate content that met those criteria. For example, for the ``Evaluations'' subcode within the ``AI Expertise'' top-level code, we select a post wherein Brendan Foody of Mercor describes evaluations as central to the advancement of AI expertise. We then recode the posts according to these subcodes, resolving differences in our application of codes through discussion.

Equipped with this initial codebook, we then review the 29 podcasts collected for this study, assigning codes and subcodes to blocks of each podcast transcript, and making note of new potential subcodes. The first and second author first code three podcasts, before meeting and comparing codes, and revising the codebook to include three additional subcodes. The first two authors then review half of the remaining transcripts, before meeting again to discuss potential new codes. This results in adding three new subcodes and merging two existing subcodes. The first two authors then repeat this process for the remaining transcripts, finalizing the codebook. Before generating themes, the authors revisit previously coded X posts and podcasts, adding or merging subcodes where appropriate.

Finally, the authors hold a meeting to generate themes that answer the study's RQs, using an affinity diagramming approach \cite{lucero2015using} to group observations within each deductive code (corresponding to a research question), extracting representative quotes from podcasts and identifying informative X posts. The authors generate a total of nine themes, including three themes each to answer the AI Expertise RQ, the Human Expertise RQ, and the Institutional Expertise RQ. These themes accordingly form the content of the Findings section of this research.

\section{Findings}

We divide our findings into three sections corresponding to our three RQs, beginning with how the firms present AI expertise; then moving to human expertise; and finally turning to institutional expertise. Throughout the findings, we refer to the podcast transcript IDs included in Table \ref{tab:podcasts} (T1, T2, T...).

\subsection{AI Expertise}

\subsubsection{\textbf{AI Expertise as Cheap}}

The data annotation firms suggest that AI expertise must be cheap, meaning both economically valuable and less expensive to produce than human expertise. In many cases, the firms discuss AI expertise in terms of return on investment, which should equal what one would expect when engaging the services of a human expert, and ideally substantially exceed this. Jonathan Siddharth of Turing describes this vision on Sourcery (T28), noting the potential for AI expertise to reshape ``\$30 trillion of knowledge work that's about to be automated.'' On TechCrunch Disrupt (T23), Brendan Foody of Mercor explains the economic logic of AI expertise facing the data annotation firms, stating that ``ultimately what they're looking at is, for every dollar they spend on us, for every hour that we're able to deliver of people working to create datasets that improve models, what's the ROI on that?''

The centrality of ROI results in firms like Scale AI and Mercor authoring or endorsing evaluations for measuring the real-world value of the expertise, providing not only a measure of AI progress but a demonstration of the value of the data extracted in the expert gig economy. One such evaluation is GDPVal, a benchmark reposted on X on September 25, 2025 by Foody and authored by OpenAI, which assesses whether agents can perform the work of industry experts with an average of 14 years of experience \cite{patwardhan2025gdpval}. Another is APEX, a benchmark authored by Mercor that purports to measure the ability for AI to automate highly valuable industries \cite{vidgen2025aiproductivityindexapex}. In an October 2, 2025 post introducing this benchmark, Foody compares AI to a PhD holder and suggests that models are ready for employment. At TechCrunch Disrupt (T23), Foody frames the post-training phase in which an AI agent learns directly from expert annotations as, ideally, a one-time investment, resulting in efficiency similar to how ``software is a fixed cost investment where you do it once and then you can run it many times.'' Despite the expectation of abundant and inexpensive expert knowledge and capability, though, expertise remains expensive to obtain, and the data annotation firms report industry-standard or better hourly rates of pay to their experts. Foody notes (T23) that Mercor's ``average pay rate right now is \$95 an hour,'' while Garrett Lord of Handshake AI says on the Grit podcast (T13) that experts on his platform receive on average ``over \$100 an hour, like \$125 an hour.'' These rates indicate the expectation that expert labor represents a one-time or few-time cost that will result in inexpensive AI expertise that can generate a return on investment.

Situating AI expertise relative to ROI demonstrates how the business model of the AI labs and the data annotation firms defines the way expertise itself comes to be understood. Similar to the way artificial general intelligence is shaped by analogy to societal undertakings like the Manhattan Project \cite{friederich2025against,katzke2024manhattan}, which demand massive public funding and support, AI expertise is shaped by an economic logic wherein cost matters as much as quality.

\subsubsection{\textbf{AI Expertise as in Need of Recontextualization}}

The data annotation firms present AI expertise as in need of recontextualization. To be useful in the real world, AI expertise should be engineered and tested in conditions that more closely mimic the real world. This means operating with access to the numerous tools and programs to which a human would have access, and dealing with the volume of information from various sources a human would encounter.

The CEOs often lament the failure of AI agents to reason about information that requires contextual and subjective judgment. Edwin Chen of Surge AI describes this on Lenny's Podcast (T4), referring to the International Math Olympiad (IMO), a highly competitive mathematics competition for high school students with problems now used to benchmark mathematical ability in LLMs \cite{luong2025imo}, to note that ``it's kind of crazy that these models can win IMO gold medals, but they still have trouble parsing PDFs. And that's because even though IMO gold medals seem hard to the average person\dots they have this notion of objectivity.'' The problem is amplified when a model must chain multiple actions together in what is sometimes referred to as a trajectory. In describing the kinds of data that Handshake hopes to collect on Lenny's podcast (T16), Lord specifically mentions trajectories and other long-horizon demonstrations that involve ``stitching together step by step instruction following,'' allowing for sustained application of varied expert knowledge.

Beyond collecting more complex and realistic demonstrations, though, the CEOs also describe the need for evaluations that more faithfully assess how models perform in the real world. Foody asserts on 20VC (T19) that ``one of the largest inefficiencies in all of AI research is that the evals that people have been going on \dots are wholly disconnected from the outcomes that consumers and enterprises actually care about.'' Siddharth observes on This Week in Startups (T25) that ``one big challenge \dots that I see today in the overall evaluation and benchmarking market is that a lot of the evaluations are somewhat academic and somewhat synthetic, and don't connect to real world applications or real world use,'' noting that good evaluations satisfy three dimensions: complexity (difficulty), real world applicability, and diversity (breadth). The situation prompts Foody to write a June 30 blog on X highlighting the importance of evaluations (and environments, as discussed below) for translating AI research into real world value.

In addition to creating evaluations for assessing models, the data annotation firms also produce new environments for training AI models. Siddharth describes on The Information TV (T27) how Turing is ``creating thousands of these [reinforcement learning] gyms across different enterprise consumer use cases,'' explaining that they ``are like these mini world models for business where the gym itself is a collection of prompts and environment and verifiers.'' He also describes these environments on 20VC (T29) as ``a mini world model with clones of these applications that are created with a fake database, with synthetic data, and the prompt might be, hey prepare for a call with this person, and then after the call is done update Salesforce.'' On Lenny's podcast (T4), Chen notes that ``we might build a world where you have a startup with Gmail messages, and Slack threads, and Jira tickets, and GitHub PRs, and a whole codebase, and then suddenly AWS goes down, and Slack goes down, and so okay, model, what do you do?'' The firms view environments as a more realistic platform for agents, affording them access to tools and information to carry out goals.

Though it may at first appear only tangentially related to the core business of the data annotation firms, creating new evaluations and environments allows the firms to influence the direction of the AI industry and drive demand for their services. Evaluations provide a measurable target that might be attained by purchasing the appropriate data, while environments provide evidence that AI expertise has not yet reached the level of fidelity expected for real-world settings, as discussed below.

\subsubsection{\textbf{AI Expertise as Unrealized}}

While the data annotation firms express confidence that AI expertise will transform the economy, they also agree that such expertise had not been sufficiently realized, and that the need for expert data will persist for years at a minimum. The CEOs sometimes face questions about whether the creation of durable AI expertise will render their businesses unnecessary, prompting discussions of both superintelligence (greater-than-human capabilities in a specific domain) and artificial general intelligence, or AGI. Chen, on Lenny's podcast (T4), contextualizes Surge AI's business model with respect to AGI, noting that ``by definition, if we haven't reached AGI yet, then there's more for the models to learn from.'' He further notes on No Priors (T2) that ``at the end of the day, you just want as much diversity and richness as you can get \dots I think there's almost an unlimited ceiling here.''

The notion of AI expertise being unrealized because the data on which it can be trained is close to unlimited also shows up in the CEOs' descriptions of demand for expert data. On DB Live (T12), Lord describes ``insatiable demand for reasoning data right now,'' and on Lenny's podcast (T16), he notes that ``in this market there's essentially like unlimited demand. Like if you can produce high quality volumes of data, you most likely will be able to sell whatever you produce.'' Foody says on 20VC (T19) that ``the total addressable market is limited by the amount of things that humans are better at than models \dots so long as there's things that the human is able to do that the model's not able to do, and we want those capabilities in the model \dots we need humans that help to create those verifiers.'' Jason Droege of Scale AI notes on TBPN (T6) that ``as we generate more and more data, the models are serving customers. Those customers have higher and higher demands of what the model should do, which then drives demand for more data, and the data changes,'' creating a feedback loop of escalating requirements that's actually driven by clients and end users of generative AI systems.

Some of the CEOs indicate that actually realizing AI expertise across the many sectors of the economy will require disruptive change. On Lenny's podcast (T20), Foody suggests that ``it's highly likely that the entire economy will become an RL [reinforcement learning] environment machine, building out all of these worlds and contexts.'' Siddharth describes a similarly totalizing vision for AI expertise on 20VC (T29), saying ``imagine this four-dimensional matrix where the first dimension is every industry, like financial services, retail, healthcare\dots the second dimension could be every function, software engineering, marketing, sales, finance, etc. The third dimension could be every role in that org chart\dots The fourth dimension is a workflow that a human goes through in that role\dots we are creating RL environments for every workflow, for every role, in every function, in every industry.''

The seemingly limitless array of human expert domains offers evidence, in this telling, of the ongoing need for the data annotation firms and the expert gig economy. Unlike many accounts of artificial general intelligence which cast it as potentially capable of automating almost all knowledge work overnight \cite{mclean2023risks,bubeck2023sparks}, the AI expertise envisioned by the firms is bound to specific demonstrations of human capability, and driven by the need to saturate every expert domain of sufficient economic value to warrant the cost.

\subsection{Human Expertise}

\subsubsection{\textbf{Human Expertise as Decoupled from the Expert}}

The data annotation firms envision human expertise as decoupled from the individual expert. As noted previously, AI expertise is presented as decontextualized from the real world and in need of recontextualization via environments and evaluations. By the same token, human expertise is envisioned as a body of ability to be extracted from the expert through these processes and imbued into AI.

The language of extraction and distillation pervades the CEOs' descriptions of the expert gig economy. On B2BaCEO (T26), Siddharth states that ``you want to distill proprietary knowledge trapped in the minds of humans that work in these companies, and distill that into LLMs''; on This Week in Startups (T25), he similarly notes that ``we've run out of internet data, but there is lots of intelligence trapped in the minds of humans that is yet to be transferred from human minds to machine minds.'' Human expertise is thus discussed as something to be freed or transferred, decoupling it from the expert. Lord describes on Lenny's podcast (T16) a common strategy for accomplishing this, by following experts ``talking over what they're doing, actually following and screen recording where their mouse is going, how they're problem solving, when they run into a roadblock, what do they do.'' Similarly, Chen describes the systems used by Surge AI to enable expertise extraction on Lenny's podcast (T4), saying ``we essentially gather thousands of signals about everything that you're doing when you're working on the platform. So we are looking at your keyboard strokes. We are looking how fast you answer things.'' Far from just discrete labeling outputs or written rubrics produced by experts, the expert gig economy employs a surveillance infrastructure to decompose expertise into signals that can be captured and analyzed independently of the expert. This enables what Chen describes on No Priors (T2) as ``technology that we're using to measure the quality that our workers or annotators are generating\dots we basically have an ML team that builds a lot of these algorithms to measure all of this.''

The professionalization of the annotation setting through which expertise is extracted (recording decisions, trajectories, and evaluations in a machine-legible way) also divorces the human expert from the professional \textit{environment} in which human expertise is created and maintained, again decoupling expertise from the life of the expert to transform their professional practice into data. This also presents a tension for the data annotation firms, as the extraction of expertise nonetheless relies on experts continuing to work in their fields. Siddharth acknowledges on This Week in Startups (T25) that ``you ideally need these humans part-time, because the way these humans are good at the job of training the models, is because they are also good at their day job, which keeps their skills sharp.'' Concerns about expert deskilling \cite{sambasivan2022deskilling,natali2025ai}, then, could be a concern not only experts but even for the organizations extracting their labor.

\subsubsection{\textbf{Human Expertise as Relative to AI Expertise}}

The data annotation firms describe human expertise as defined relative to the knowledge and capabilities of an AI model or agent, such that human expertise can be characterized as the ability to improve, break, or otherwise beneficially alter a model. Lord expresses this most directly, across multiple podcast appearances. On TBPN (T9), he asks ``what better way to articulate your skill than actually proving it by being able to break the model, or by being able to provide the model feedback?'' He further explains on DB Live (T12) that ``what these frontier labs need to improve their models is experts, and experts that really are even better than the model, to actually break down and explain why a model's getting something wrong, and what the right answer is.'' On Lenny's podcast (T16), he notes ``the models have gotten so good that the generalists are no longer needed. What they really need is experts\dots it's pretty tough for the average person to break a model and get an incorrect response. But if you're a PhD in physics, you can go in, in multiple subdomains of physics, and prove where the model's actually breaking.'' Human expertise is valued precisely in domains where AI expertise hasn't approximated it.

Siddharth similarly describes human expertise as relative to AI on Sourcery (T28), saying ``the data that the models need now needs to be hard, that is, model breaking. It has to literally break the models, so you need humans who are smarter than the models.'' On This Week in Startups (T25), he further notes that ``two years ago a very low-skilled contractor was capable of creating tokens that could have advanced the model. Now, because the floor has gone up, now you need PhDs from Stanford, Berkeley, MIT to figure out where the models break.'' Siddharth also states (T25) that ``with the models advancing like this\dots you don't just need Iron Man, you need the Avengers\dots frontier models need frontier data. Frontier data needs a team of frontier humans,'' positioning human expertise as a necessary input to frontier AI, with value defined by that relationship. Droege describes the ubiquity of this position on TBPN (T6), noting ``everyone in the data business is in the expert business, because that's what the models need, that's what they demand.'' 

The industry's description of expertise inverts traditional ways of comparing human and AI knowledge and capability. When AI can approximate human performance on shared standards, the data annotation firms suggest that human expertise can be best measured by comparison to what has already been encoded in a machine. In this way, the present form of the product, the AI model or agent, determines what can be considered expertise among the humans who use and produce that product. Human expertise necessarily becomes more fragile and more temporally contingent in such an environment, as further described below.

\subsubsection{\textbf{Human Expertise as Elastic}}

The data annotation firms describe human expertise as elastic, meaning that it is made available and extracted on demand, depending on the needs of the AI labs. Because human expertise has value in the expert gig economy relative to the capabilities of AI, the need for different forms of expertise changes depending on the performance of the newest models. Such model-driven demand on the part of the frontier labs creates temporal cycles in expertise needs. Lord notes on Lenny's podcast (T16) that ``each one of the research teams inside of the labs have different use cases \dots they have a hypothesis around how to improve the model. They're trying to collect small pieces of data to see if that hypothesis works out. If that hypothesis is proving true, then they expand the overall collection of the data in that effort.'' Siddharth discusses the need for elasticity as responding to lab needs that vary by quarter, noting on B2BaCEO (T26) that ``you need a platform that can spin up and down, because the needs of the labs change quarter over quarter.'' On the more extreme end of the spectrum, Chen speaks on 20VC (T1) about responding to demand, saying ``customers\dots will literally call me at 2am, 3am, and they'll be like, hey, our models are freaking out. I need a bunch of data to fix it by 6am. Can you do it? \dots Yeah, we can deliver 10,000 data points to you in the next few hours.'' Such responsiveness requires experts as an elastic resource that can scale up almost instantaneously.

For experts, elasticity provides opportunities in the form of, as Lord states directly on TBPN (T9), ``a gig job.'' He further describes the arrangement on The Information TV (T15), saying ``you might be able to come in for five, ten, fifteen hours if you're a really great accountant and actually produce data that is improving models.'' Human expertise is thus commodified in the expert gig economy by dividing it into small, flexible, on-demand time units. On Lenny's Podcast (T16), Lord reasserts the benefits of this arrangement, noting that experts, including those still in training, such as Ph.D. programs, ``can make like \$100, \$150, \$200 an hour in their area, in their field of expertise \dots you can actually make \$150 an hour breaking the latest models.'' Premium pay can make the work appealing, but the structure remains centered around just-in-time, hourly, on-demand work. However well paid, experts must contend with opportunities dependent less on their own abilities than on the changing needs of the AI labs.

\subsection{Institutional Expertise}

\subsubsection{\textbf{Institutional Expertise as in Need of Liberation}}

The data annotation firms suggest that part of their value lies in the ability to democratize institutional expertise, freeing it from inflexible ``legacy'' businesses. Foody explains this perspective most clearly on TechCrunch Disrupt (T23), saying that ``the reason that the labs need us is that their customers don't want to give them data to automate large portions of their value chains. They need to hire contractors that previously worked at those companies that are willing to understand those workflows, and are willing to train models how to automate them.'' When asked about whether Mercor hires people still working at these organizations, he clarifies that ``for the most part it would be people that previously worked at those companies,'' and notes that ``we have contracts with all the experts that say they're not supposed to'' bring data from their old company. Describing what he thinks about who owns the institutional knowledge built during employment, Foody responds, ``does the knowledge that lives in a former Goldman banker's head belong to Goldman or does it belong to that individual? I think that it belongs to the individual.'' The debate over data ownership continues to play out in the discourse over how AI data is obtained and repurposed \cite{cooper2025extracting,ahmed2026extracting,bender2021dangers,jiang2023ai}. This question now appears to have come for the specialized, non-public knowledge developed, perhaps at great cost, by institutions that employ experts.

Siddharth expresses a more optimistic perspective on the evolution of institutional expertise. Rather than focusing on the inflexibility or guardedness of present institutions, he notes on 20VC (T29) that ``for a lot of ideas, founders are intelligence constrained\dots for example, if you pick a therapist who wants to start a mental health startup, today that founder would have to raise at least a few hundred K, if not a few million, to recruit some software engineers, maybe a marketing person or a growth person, maybe a product manager.'' Instead, he notes that in the future, these roles can be carried out by inexpensive, custom-trained GPT models. For Siddharth, institutional expertise will be liberated by cheap AI expertise, but in a way that empowers individuals to create new institutions. As he goes on to say (T29), ``a million flowers will bloom. Lots and lots of nontechnical founders will start companies. We'll see a broader distribution of founders than just those who live in London or Palo Alto \dots which I think is wonderful for the world.''

Whether that future plays out, though, may depend on how sustainable the data annotation firms' position is in relation to the businesses they expect to be transformed by cheap AI expertise. Their status as intermediaries provides them with a notable advantage in that they can offer gig work to experts formerly employed at traditional institutions, and direct the data harvested to the AI labs. Rather than attempting to purchase proprietary data from an organization that does not wish to sell it, the AI lab can work with individuals with institutional knowledge to indirectly approximate that data. However, Foody's appearance at TechCrunch Disrupt surfaced the potential for rising tensions between businesses and the data annotation firms seeking to replicate institutional expertise, suggesting that the firms' position might not be as inevitable or as unassailable as they imply.

\subsubsection{\textbf{Institutional Expertise as Performative}}

The data annotation firms suggest that institutional measures for signaling expertise are obsolete and primarily performative in the time of AI, and that the annotation firms themselves are well-positioned to create new approaches to measuring and communicating expertise. Though they also rely on university credentials within their expert networks to communicate their own credibility, many CEOs express skepticism about the power of a university degree, as well as the ability of HR departments at many institutions to respond to increases in job submission materials due to AI.

Chen provides an assessment of advanced degree holders on 20VC (T1), saying ``what people underestimate is that having a PhD isn't enough. Like a lot of PhDs, they just aren't good at this type of work\dots I think 80\% of the computer science PhDs I know, they write shitty code because they're only good at math and algorithms. And then think about people like Ernest Hemingway. He didn't have a PhD. I don't think he even went to college.'' Foody shares his own personal experience rejecting institutional credentials on TechCrunch Disrupt (T23), noting that ``I told my parents every semester that I was going to drop out of college, because initially I didn't want to go in the first place. And then by the time that I really dropped out, I just didn't go to finals, my final semester.''

Of course, despite their evident skepticism, most of the firms tout their ability to recruit advanced degree holders to their platform. Siddharth, for example, notes on This Week in Startups (T25) that Turing is ``hiring PhDs across the board in physics, chemistry, math, biology, at the level of granularity of somebody who's an expert in dark matter,'' while Chen asserts on 20VC (T1) that ``if you think of all the PhDs, even at Google or Meta or Microsoft, we have way more than all of them combined.'' Though the CEOs may see the credentialing systems offered by longstanding institutions as primarily performative and outmoded, they still rely on them, at least for the time being, for establishing the credibility of their own organizations.

Of the data annotation firms, Handshake has the longest relationship with universities in particular, having partnered with universities to provide a portal for job seekers for more than a decade, and still viewing them as a primary client base. It's perhaps unsurprising, then, that Lord offers the most coherent suggestion for a replacement or supplement to existing models of institutional credentialing. As discussed above, Lord and other CEOs believe that if an individual can break a model in a way that could ultimately improve it, the individual has some claim to expertise that has economic value. The logical outcome of this, proposed by Lord, is to offer these individuals not only financial compensation but also credentials that communicate their expertise to employers. As he notes on TBPN (T9), ``we envision a world where you get badges on your profile, and there's leaderboards by school \dots we believe that we can help you get more jobs with the million employers in the [Handshake] network, help you build your professional reputation and articulate your skills.'' For Lord, the credential would signal not only that the individual has some form of expertise, but also that they understand how to work with AI, and how to impart their expertise to AI. Though Lord's vision appears largely concerned with credentialing individuals, he mentions on Let There Be Light (T10) that there could be ``leaderboards by schools'' (plural), which may suggest a ranking system that evaluates academic institutions themselves on their ability to produce strong data annotators. Though such a system may at first be treated as a competitive diversion, it could eventually be perceived as a real signal of institutional contributions to AI, and potentially inform scores assigned by other educational ratings institutions, especially if student career outcomes were to depend in part, as Lord suggests, on credentials earned by participating in the data annotation economy.

\subsubsection{\textbf{Institutional Expertise as Inverted by Augmentation}}

The data annotation firms suggest that institutional expertise will be inverted by a generation of young workers who can draw on AI expertise, reshaping the creation and transfer of knowledge and power within institutions. These suggestions often respond to the concern that cheap AI expertise produced by the expert gig economy will harm the entry-level job market, primarily affecting young people. The CEOs typically answer this concern with the assertion that greater familiarity with AI will instead enable young people to do more overall and to do it more quickly, in contrast with more experienced employees who might be reluctant to adopt. Lord describes his vision of AI use as akin to putting on an ``Iron Man'' suit on DB Live (T12), saying ``I like to think about it as, how can you build an Iron Man suit around yourself to exponentially increase the productivity of what you can accomplish inside the workplace?'' On The Information TV (T15), he says ``you can just bolt on an Iron Man [suit] to most roles, right? Ten years ago, maybe fifteen years ago, you added Google search as a skill on your resume, right? and people that knew how to use Google search significantly outperformed their peers.'' For Lord, it is young people who will be ready for this, as he notes on The Peel (T14) that ``young people have an opportunity [to be] \dots digitally enabled, Iron Man equipped, AI native.'' On The Information TV (T15), he says ``young people will really benefit from this wave. These students and young professionals are AI native\dots we just had an intern who put up a pull request on his first day\dots young, tech-native employees will be the ones putting up most of the points.''

The idea of a young person with technical skills disrupting an organization or industry is not novel. However, the transformation described by the CEOs is striking in its scope, applicable not just to singular software engineers or tech company founders, but to any individual with access to an LLM. Drawing on the idea of the ``10x'' software engineer, who can produce much more useful code than their peers, Chen states on 20VC (T1) that ``there are some people who simply have ideas that other people can't think of\dots two or three X is often an underestimate\dots you get to 100.'' Lord similarly describes AI as a productivity multiplier on the The Information TV (T15), noting that ``I think LLMs and AI is like a factor of a hundred more effective than Google search in terms of the productivity it adds to each knowledge worker.''  This view of enhanced individual productivity also leads to the CEOs envisioning AI-augmented individuals as capable of doing more than one job, or even creating and leading multiple organizations. Siddharth suggests on 20VC (T29) that if ``I'm 100x more productive, maybe I'm able to run a hundred companies \dots I think every human will just be so much more leveraged\dots we are accustomed to the idea of one person doing one job. But people could be doing multiple jobs at the same time. People could be running different companies at the same time.'' 

Though the idea of a vastly more productive workforce centered around opportunities for young people sounds optimistic, the presentation of young people as AI-augmented superheroes also inverts expectations for most entry-level knowledge work jobs. Rather than building knowledge and skills to prepare one for a long career in one's industry, the AI-augmented employee already has access to what's needed for success; the only limitation is the willingness to integrate AI sufficiently into one's professional life to be fully effective. Such presentations also have implications for experienced employees who might typically serve as gatekeepers and disseminators of institutional knowledge. The expertise of more experienced peers and managers is rendered in some ways suspect, in that best practices learned in an economy before AI might prevent one from moving fast enough to contribute to the extent expected. Were AI (or more likely AI-augmented competition) to marginalize the value of their expertise, these individuals may find their status and employment threatened, and may be forced to adopt AI themselves to compete.

\section{Discussion}

The expert gig economy appears, for the moment, to differ from past gig economies. Its workers are better credentialed, better paid, and more likely to live in wealthy, western countries than AI alignment gig workers \cite{perrigo2023openai}. In public communication, the data annotation firms speak highly of the expert workers, touting the subject expertise they bring to the firm's network, departing from research finding that data requesters fail to appreciate data workers \cite{rothschild2024problems,sambasivan2022deskilling}.

Yet there are hints, already, of the potential precarity of their situation. Foody describes Mercor's annotation workforce as following a power law, wherein the top 20\% contribute much more than the bottom 80\%, suggesting that many workers may experience algorithmic insecurity due to the less consistent opportunities afforded to them. Moreover, the data annotation firms seek to shape the market for expert data by producing their own evaluations and environments, quantifying the impact of annotations on model performance and driving demand accordingly. Expert work is also increasingly quantified using advanced surveillance technologies, suggesting the potential for experts' agency to be reduced as they have less control over what data is collected about them and the decisions made in light of it. The firms' strategies form a foundation that affords them control of the market and the workers, suggesting the potential for a coming algorithmic insecurity, similar to the experiences of workers in prior gig economies.

Moreover, the agentic technologies developed by the expert gig economy increasingly see deployment in the traditional workplaces of experts, displacing systems of expert achievement and credentialing, and pushing established experts to adopt new AI tools to avoid displacement. Even those experts who do not participate in the gig economy may thus experience increased precarity as a result of its objectives. This adds complexity to accounts of ``blended'' human-AI workplaces \cite{constantinides2025future}, as the agentic models with which workers must interact originate from a market whose objective is return on investment through automation. The arrival of agents may also portend an increase in workplace surveillance technologies in traditional jobs, as organizations seek to extract signal to mirror the expertise of workers. The future of expertise, AI, Human, and Institutional, is linked to the expert gig economy. We thus close this work by questioning the vision for expertise put forth by the industry, foregrounding concern for human experts and institutions.

\subsection{Implications and Provocations}

We consider the implications of our work for AI, human, and institutional expertise, posing provocations to prompt reflection on the future of expertise and the expert gig economy.

\subsubsection{\textbf{AI Expertise}}

Though the data annotation firms focus on the potential benefits of cheap AI expertise (such as access to programming skill, or design abilities), we ask whether more attention should also be paid to the potential societal costs of this technology. The product currently made available by the AI labs suggests not only the affordable or inexpensive meanings of ``cheap'' but also the word's derivative and lower-quality connotations, as evident in many audits and evaluations of AI systems that highlight meaningful differences between AI and human expert performance in areas like therapy \cite{wang2025feel,moore2025expressing} and law \cite{magesh2025hallucination,akter2025comprehensive}. While society awaits the arrival of reliable AI expertise, intermediate technologies are often advertised as domain specialized, only to be later demonstrated to be unfit for purpose \cite{raji2022fallacy,rauh2024gaps}. 

We also ask whether the project of producing AI expertise might operate in ways that offer more clear benefits to the public. By many accounts, the data annotation firms are efficient organizations, capable of delivering massive amounts of data to support model upgrades in a short amount of time. However, both the models created from their work and the datasets they harvest from experts remain, for the most part, proprietary, with the exception of the evaluations created by the firms, which afford an opportunity to exert influence over the data market. Yet in a context where success might entail a radical reshaping of the expert professions, we ask whether there are alternatives to this approach that could yield more open, public artifacts, as well as opportunities for public input into the process of producing AI expertise. Policy is one way of moving toward action that could achieve this. Another might be the creation of new open research acceleration programs, similar to those envisioned for social media \cite{bail2023we}.

\begin{tcolorbox}[colback=gray!10, colframe=gray!50, boxrule=0.5pt, arc=3pt, left=5pt, right=5pt, top=5pt, bottom=5pt]
\noindent \textbf{AI Expertise Provocations:}
 \begin{enumerate}
     \item What are the societal costs of developing cheap AI expertise?
     \item How might we envision a more open approach to producing AI expertise that gives back to the public?
     \item How might policy or public accelerators afford the public control over the development of AI expertise?
 \end{enumerate}
\end{tcolorbox}

\subsubsection{\textbf{Human Expertise}}

The perspectives of the expert data annotation industry raise questions about how human expertise should be cultivated and valued. To value human knowledge and capability only insofar as it cannot be replicated in AI rings hollow; yet the potential for human expertise to be economically devalued by cheap AI expertise suggests the need for a deeper and more contextual understanding of what makes human expertise valuable and desirable. This may be easier for some domains, especially those that involve more human interaction, which has inherent value for most people \cite{brinkmann2023machine}. Moreover, though expert knowledge workers have advantages that workers in previous gig economies did not (higher pay, greater options), history shows that these benefits may not last as the power of the new platforms solidifies \cite{bonini2024algorithms}. Experts might thus look to the experiences of gig economy workers past, then, who are experts in their own right on the precarity and insecurity associated with platform work \cite{wood2021platform}, to guide the path toward developing protections in this workplace. Finally, we observe the need to consider the impact and the options for workers to respond to new, more advanced surveillance technologies described by the data annotation firms. Such technologies may be used not only in data annotation, but in traditional jobs as well, as organizations harvest data from employees in service of automation.

\begin{tcolorbox}[colback=gray!10, colframe=gray!50, boxrule=0.5pt, arc=3pt, left=5pt, right=5pt, top=5pt, bottom=5pt]
\noindent \textbf{Human Expertise Provocations:}
 \begin{enumerate}
     \item How should society value expert knowledge or ability that has been automated in part or in full by AI?
     \item How can the experiences of workers in other gig economies inform approaches to worker protections in the market for expert data annotation?
     \item How can knowledge workers respond to the deployment of new surveillance infrastructure intended to capture human expertise? 
 \end{enumerate}
\end{tcolorbox}

\subsubsection{\textbf{Institutional Expertise}}

Perhaps the most central question facing the data annotation firms is the one put to Foody at TechCrunch Disrupt: to whom does institutional expertise belong? Foody sided with the individual over the institution, implying that the individual should have the option to sell their expertise in the expert gig economy. Yet this point may not be settled, particularly if the creation of AI expertise undermines the business model of an institution. As of January 2026, stories in Wired \cite{knight2026handshake} and TechCrunch \cite{ha2026handshake} described Handshake AI as requesting work product from experts' current and former jobs as part of the annotation process, evidently at the request of OpenAI. This suggests that mere approximation of institutional knowledge may not be enough, and hints at potential for legal disputes. Such disputes would mirror earlier actions arising from the use of copyrighted materials in AI pretraining \cite{grynbaum2023times,samuelson2023ongoing}.

Institutions, and universities in particular, may also need to consider the utility of their credentialing systems in the time of AI. It is not clear that AI data annotation can serve as a reliable signal of expertise as suggested by Lord, in part because many forms of expertise are not so easily reduced to demonstrations that can be reproduced in a user interface by a machine \cite{diaz2024makes}. Yet the firms have identified a real problem in pointing out that a deluge of AI-generated resumes, cover letters, and college application packets has overwhelmed traditional evaluation systems and prompted many individuals to adopt AI tools to keep pace \cite{ebanks2025hiring,brachman2025current}. Finally, we note the importance of institutions for legitimizing expertise, offering norms and social contexts that define what it means to be an expert \cite{carr2010enactments,pakarinen2025relational,mieg2006social}. We thus ask, in closing, how institutions can approach this role for a public that has direct access to (real or illusory) cheap AI expertise, without the training or credentialing that in the past may have been needed to earn that expertise.

\begin{tcolorbox}[colback=gray!10, colframe=gray!50, boxrule=0.5pt, arc=3pt, left=5pt, right=5pt, top=5pt, bottom=5pt]
\noindent \textbf{Institutional Expertise Provocations:}
 \begin{enumerate}
     \item To whom does institutional expertise belong?
     \item How can institutional credentials adjust for an AI-centered information economy?
     \item How can institutions continue to play the social role of legitimizing human expertise for a society that has easy access to cheap AI expertise?
 \end{enumerate}
\end{tcolorbox}

\subsection{A Starting Point for Critical Conversations}

We view this work as the starting point for critical conversations about the future of expert work and the future of AI. The process of creating, deploying, and integrating LLMs as workplace tools involves many stakeholders, who shape both the needs that drive the gig economy and the character of LLM usage in knowledge work across economic sectors. We thus suggest several lenses that future work might adopt in building on this research.

The first lens draws on value-sensitive design (VSD), a methodology that focuses on identifying stakeholders, eliciting their values, and envisioning long-term futures that accord with those values \cite{friedman1996value,friedman2017survey,alshehri2020scenario}. Our work suggests stakeholders ranging from model trainers at industry labs to doctoral students annotating data part time to ``legacy'' businesses concerned with the value of their institutional expertise. VSD could be used to map these stakeholders, place their values in conversation, and offer a concrete view on the long-term implications of an expert data gig economy.

The second lens might employ a social constructionist approach to probe the meaning of AI ``expertise'' in particular. Social constructionist perspectives consider expertise as a product of the social context in which the expert operates \cite{agnew199710}, such that an expert is afforded a position of privilege due to their ability to offer guidance that meets the needs of some population in society, often after a period of rigorous training in a specific context \cite{grundmann2017problem}. Future work might probe, then, whether the rigorous training and interpersonal context by which expertise is acquired have value that surpasses the approximation offered by recontextualized AI expertise, both for clients of experts and for society more broadly. 

The third lens draws on knowledge management and documentation practices in computer-supported cooperative work, which could study how needs for expert demonstrations and associated practices of organizing expert data change over time. Past work establishes that knowledge and expertise are not productively regarded as static or fixed, but require knowledge management practices to revise and expand bases of knowledge as needed \cite{ackerman2013sharing}. Our research identifies a tension that might be studied through this lens, namely whether and in what ways modular, extracted expertise can approximate the reusability of software, especially when the firms themselves expect ongoing demand for large volumes of expert data that will continue for years at a minimum. Given that this demand is driven at least in part by advances in the complexity and fidelity of the environments in which the models train, a knowledge management approach might offer a way of understanding how expertise is being captured, stored, and reused across this rapidly changing sociotechnical setting.

Finally, the fourth lens might further consider how the gig economy intersects with work on workplace surveillance, including questions of how work environments are reshaped by technologies designed to capture expertise \cite{levy2022data}, and the potential for equity-focused alternatives to extractive perspectives \cite{breznitz2025equity}. The widespread deployment of surveillance technologies that document the activities of expert annotators may affect all knowledge work, both indirectly by affording the creation of AI expertise with the potential to devalue human expertise, and directly by providing a model for broader application of these technologies in all jobs. As observed in much prior work on justice in computing, concern for the people most vulnerable to the adverse effects of a technology is often ultimately concern for society as a whole \cite{chordia2025building,chordia2024social}, as deployment among vulnerable populations may serve as a stepping stone to application at larger scale \cite{eubanks2018automating}.

\subsection{Limitations}

The chief limitation of our work lies in its focus on the public communication of the data annotation firms, and particularly on podcasts that speak primarily to investors. While this captures an important perspective on the future of expertise, it also limits our analysis to 1) only communication from the companies \textit{themselves}, that they chose to disseminate online; and 2) primarily communication meant to motivate financial interest in the companies. In both cases, we likely capture data that reflects the intentions of the firms as much as the reality of data work in these organizations. Thus, our research could benefit from complementary studies that foreground the perspectives of other organizations in the LLM ecosystem with respect to the role of the data annotation firms; and, more critically, it could benefit from complementary work that foregrounds the perspectives of expert data annotators themselves. It could also benefit from a comparison of the public communication of the data annotation CEOs against that of other tech CEOs, including those leading AI and machine learning companies, to better situate the perspectives of the data annotation firms within the industry.

Our work is also limited in that we analyze data from five firms during a seven month timeframe in 2025. While this timeframe affords us a substantial quantity of data for qualitative analysis from an eventful time in the industry, the need for diverse forms of data to train AI continues to evolve quickly, and future work will be needed to consider how the expert gig economy changes, and ideas about the future of expertise change with it. Moreover, our selection of the five organizations relies on 1) their identification by major media outlets as central organizations in the data annotation space; and 2) availability of public data from the firm for analysis. An alternative approach might have intentionally selected organizations that vary across financial profiles (\textit{e.g.}, both large and small companies) or based on the specific forms of expert data for which the organization is most known. Finally, our work focuses primarily on expert data annotation in a western, primarily American context. Future work might accordingly study the character of expertise and the expert gig economy across cultural contexts.

\section{Conclusion}

This research describes the vision of the expert data annotation industry for the future of expertise, discussing expertise in terms of three complementary aspects of AI, Human, and Institutional expertise. We show that the industry views AI expertise as cheap; human expertise as extractable and relative to AI; and institutional expertise as in need of liberation and inversion. We then offer provocations for each form of expertise, intending to prompt reflection on the implications of the expert gig economy, and alternatives to the vision put forth by the data annotation firms. We expect our work will serve as a useful guide for future work studying the norms of this rapidly evolving industry.

\section*{Statement on the Use of Generative AI}

We used Claude Opus to review and critique near-final drafts of this manuscript, after completing our thematic analysis. In some cases, we revised the paper to make our arguments stronger, incorporating more evidence (primarily by adding additional quotes) where the model identified weaknesses in the draft.

\bibliographystyle{ACM-Reference-Format}
\bibliography{references}

\end{document}